\documentclass[12pt]{article}
\usepackage{subeqnarray}
\usepackage[title]{appendix}
\usepackage{epsfig,amsmath,amssymb,mathtools}
\usepackage{enumitem,setspace}
\usepackage{mathrsfs}
\usepackage{empheq}
\usepackage{soul}
\usepackage{centernot}
\usepackage[usenames,dvipsnames]{color}
\usepackage[pagebackref=true, colorlinks=true]{hyperref}
\definecolor{redish}{rgb}{0.7,0.2,0.0}  
\definecolor{bluish}{rgb}{0.2,0.5,0.8}
\hypersetup{linkcolor=redish,          
                  citecolor=blue,        
                  filecolor=magenta,      
                  urlcolor=bluish}          

\newcommand{\nn}{\nonumber}
\def \({\left(}
\def \){\right)}
\def \[{\left[}
\def \]{\right]}
\begin{document}
\title{Logic, Philosophy and Physics: a critical commentary on the dilemma of categories}
\author{Abhishek Majhi,\\ Indian Statistical Institute,\\Plot No. 203, B. T. Road, Baranagar,\\ Kolkata 700108, West Bengal, India\\
	(abhishek.majhi@gmail.com)}%
\date{~}
\maketitle
\begin{abstract}
I provide a critical commentary regarding the attitude of the logician and the philosopher towards the physicist and physics. The commentary is intended to showcase how a general change in attitude towards making scientific inquiries can be beneficial for science as a whole. However, such a change can come at the cost of looking beyond the categories of the disciplines of logic, philosophy and physics. It is through self-inquiry that such a change is possible, along with the realization of the essence of the middle that is otherwise excluded by choice.  The logician, who generally holds a reverential attitude towards the physicist, can then actively contribute to the betterment of physics by improving the language through which the physicist expresses his experience. The philosopher, who otherwise chooses to follow the advancement of physics and gets stuck in the trap of sophistication of language, can then be of guidance to the physicist on intellectual grounds by having the physicist's experience himself. In course of this commentary, I provide a glimpse of how a truthful conversion of verbal statements to physico-mathematical expressions unravels the hitherto unrealized connection between Heisenberg uncertainty relation and Cauchy's definition of derivative that is used in physics. The commentary can be an essential reading if the reader is willing to look beyond the categories of logic, philosophy and physics by being `nobody'.
\end{abstract}
\newpage
\tableofcontents
\newpage
   	\section{Introduction}
   	When the physicist speaks of ``logic'', he does not care about being logical in the sense of the logician who delves for arithmetization of language, let alone the fact that the physicist uses the same symbol, with different senses, in the same process of reasoning while performing dimensional analysis\cite{ll}. Plethora of examples can be given, but a few should be enough to draw the attention of the reader. Einstein speaks of ``contradictions'' in the introductory part of ref.\cite{einsr} as he makes an {\it assumption} ``{\it that the definition of synchronism is free from contradictions}'', on page no. 45 of ref.\cite{principlerel}. However, he does not provide any formal analysis with the propositions, the assumptions and the postulates from which these ``contradictions'' can be understood from the logician's point of view. Although Newton axiomatizes physics in refs.\cite{principia1,principia2}, however still today the physicist never does a formal logical analysis in the sense of what the logician does e.g. see refs.\cite{church,hilbertackermann,mendelson,heijen}. Possibly due to such reasons Poincare writes, regarding the arithmetic precision of the physicist's reasoning,  ``{\it that it is precisely in the proofs of the most elementary theorems that the authors of classic treatises have displayed the least precision and rigour.}'' (see page no. 6 of ref.\cite{poincare}). Although he further points out that ``{\it they have obeyed a necessity}'', but he does not explain what the ``necessity'' is. As it appears to me, the ``necessity'' is to operate i.e. to do something of a practical essence which one can experience materially (through bodily perceptions). It is such an operational mindset that founds the basis of all efforts by the physicist\cite{logicphysics}. Indeed the term ``definition'', without which any theory of physics is crippled, is also taken in this particular sense of being operational, rather than being logical in the sense of the logician\cite{ll}. Notably, motivated by ``{\it Einstein's treatment of simultaneity}''\footnote{Regarding being ``operational'', Einstein has been motivated by Mach\cite{machmech}.}, Bridgman considers ``definition'' to be ``operational'' by writing that ``{\it the true  
	meaning of a term is to be found by observing what a man does with it, not by what he says about it. }'' (see page no. 7 of ref.\cite{logicphysics}).

   But, then why does the physicist talk about logic? What is the necessity of the axioms, the postulates, the theorems, etc. if the reliance is on being operational? It is because without ``saying'' or writing, theories can not be written and without written theories experiments have no essence. So, language and its logical structure do have an indispensable role in science.  And here comes the question that how truthfully the physicist expresses the experience. It appears to me that the physicist's expressions of experience are susceptible to very basic questions of reason\cite{ll}. 
     
    Why does he write ``zero length'', instead of writing ``negligible length'', while seeing a dot of the pencil on a paper \cite{essay}? Is it just because he wants to base his theory, on the first axiom of geometry rather than on practical experience and be logical  rather than being operational? If I consider the very basic traits of a table top experiment, why does the physicist ignore the thickness of the cross-wire while writing theories to explain the observation and measured data, in spite of being unable to dispense with that thickness in the process of measurement? 
    
    If the physicist can not take into account such simple truths of experience, then how can he write down a theory of everything\cite{hawking} or even dream of a final theory\cite{weinberg} or provide a complete guide to reality\cite{roadtoreality}?
    
    I believe, it is the inability of judging one's own truthfulness or, in other words, a lack of self-inquiry that leads the physicist to believe in achieving completeness -- a belief so strong that it makes the rational mind incapable of realizing the impossibility of the expression of the self and the essence of the middle way that provides the ground for categorization of truths e.g. see section (2) of ref.\cite{ll} for a simple demonstration with drawing. Importantly, the logician and the philosopher, who are  supposed to carry out such inquiries regarding the language of the physicist, chooses not to question, and sometimes shows a reverential attitude towards, the physicist. Indeed it is the physicist who, at times, have questioned the basic principles with which the logician operates \cite{vonneuman,finkelstein,putnam,reichenbach,dummett,quantumlogic} and have doubted, or rather dismissed altogether, the significance of the philosopher in the context of physics\cite{hawking,weinberg}.

    In what follows, I provide some comments meant for the logician and the philosopher in order to explain how the logician lacks the attitude of self-inquiry and how both choose not to question the physicist's dogma, which otherwise could have led physics through a different path of evolution. Many a times, at the cost of sophistication of language, the truthfulness of the expressions of experience, is sacrificed by the logician and the philosopher. A change in attitude by the logician and the philosopher to look beyond the categories may be beneficial to the physicist and fruitful for science as a whole. 
    
     Although the essence of the present work is best realized in relation to ref.\cite{ll}, but reading this in isolation can also be a fruitful experience.  I hope the following commentary proves to be worthy of reading to the reader who is willing to see through beyond the categories of logic, philosophy and physics by being `nobody'\cite{ll}.

     {\section{Comments for the logician}\label{logicians}
     	Tarski writes on page no. 152 of ref.\cite{tarskiundef}):\vspace{0.1cm}

     	{\small ``{\it The question how a certain concept is to be defined is correctly formulated only if a list is given of the terms by means of which the required definition is to be constructed. If the definition is to fulfill its proper task, the sense of the terms in this list must admit of no doubt.}''}\vspace{0.1cm}
     	
     	Tarski's assertion that the definition of a concept ``must admit of no doubt'' manifests the general belief of the logician that there must be completeness of reason. Such an attitude disregards that the terms of the list themselves can retain doubt i.e. the premise itself may be questionable. This can be understood by one's attitude of self-inquiry that provides the seed of doubt, by questioning the self --  do I know the meaning of each and every term? If not then how can the definition fulfill its proper task? It is true that such self-inquiry has to stop somewhere so as to have a working definition. However, then the doubt is always retained. To have this realization there is no necessity to learn the specific jargon of mathematical logic. Rather it depends on self-awareness. It is on such grounds that the logico-linguistic inquiry regarding the foundations of physics has been done in ref.\cite{ll}. The logician's lack of self-inquiry and his exclusion of the middle from consideration leads him to the belief of completeness. I provide the following commentary on Frege's views on completeness of definition, which the reader may compare with the discussion in section (2) of ref.\cite{ll} for a better understanding.

      \subsection{Frege on Definition: a metaphor and self-inquiry}\label{fregedefmet}
 In highlighting the logician's attitude to achieve completeness of logic it is worth mentioning Frege's work in ref.\cite{fregedef}.      
Frege's attitude towards achieving completeness in the definition of a concept reflects from his naming of the introductory section as ``Principle of Completeness'' on page no. 159 of ref.\cite{fregedef}. Further, he writes:\vspace{0.1cm}

{\small ``{\it A definition of a concept (of a possible predicate) must be complete; it must unambiguously determine, as regards any object, whether or not it falls under the concept (whether or not the predicate is truly assertible of it). Thus there must not be any object as regards which the definition leaves in doubt whether it falls under the concept;...}''}\vspace{0.1cm}

Frege asserts that the definition of a concept must be ``complete'' and there should be no place for ``doubt'', although he is not sure whether his judgment about the definition of a concept is decisive or not, as he continues to write: \vspace{0.1cm}

{\small ``{\it ...though for us men, with our defective knowledge, the question may not always be decidable.}'' }\vspace{0.1cm}

 This indecisiveness of Frege, if analyzed with {honesty through self-inquiry}, gets manifested in the expressions of his thoughts through a geometric metaphor as follows:\vspace{0.1cm}

{\small ``{\it We may express this metaphorically as follows: the concept must have a sharp boundary. If we represent concepts in extension by areas on a plane, this is admittedly a picture that may be used only with caution, but here it can do us good service.}''}\vspace{0.1cm}

Here, I may note that Frege  writes about ``expressing'' a metaphor. Now, any expression of a human being comes out through some mode of expression. Since Frege wants to express by drawing lines to form areas on a plane, he needs a drawing object, say, a pencil whose tip has a spread or thickness that gets manifested on paper as the thickness of a point and therefore, of any line drawn with that pencil.  I do not know what ``caution'' Frege tries to point out. However, if it is the irremovable thickness of the pencil-tip that stands for the ``caution'', then this metaphor only does limited service as far as the completeness of a concept is concerned. To explain this I need to quote Frege further as follows:\vspace{0.1cm}

{\small ``{\it To a concept without sharp boundary there would correspond an area that had not a sharp boundary-line all round, but in places just vaguely faded away into the background.}''} \vspace{0.1cm}

Here, I emphasize that the ``vaguely faded away'' portions can only be removed up to the thickness of the pencil-tip. Therefore, the notion of completeness of the definition of a concept must depend on the context as, metaphorically, the concept of the area is precise only up to the thickness of the pencil-tip. If the pencil-tip does not have a thickness then the drawn line also has no thickness. Such line can not be seen so that an expression through a drawing can be given.  It is such a directly observable truth that Frege ignores while advocating his ``Principle of Completeness''. If I write -- ``up to now no one has ever seen a line with no thickness'' -- then this statement is a direct derivative of what Frege himself writes on page no. 11 of ref.\cite{frege1}:\vspace{0.1cm}

{\small ``{\it ..... for up to now no one.... has ever seen or touched 0 pebbles.}''}\vspace{0.1cm} 

I can safely conclude that completeness comes with the ignorance of incompleteness. In other words, doubt can not be removed completely as the truth depends on the context and hence, is only relational and not absolute. 

Now, Frege writes further: \vspace{0.1cm} 

{\small ``{\it This would not really be an area at all; and likewise a concept that is not sharply defined is wrongly termed a concept.  Such quasi-conceptual constructions cannot be recognized as concepts by logic; it is impossible to lay down precise laws for them. The law of excluded middle is really just another form of the requirement that the concept should have a sharp boundary.}''}\vspace{0.1cm}

Corresponding to this passage I may write as follows. The  area with a sharp boundary that Frege wants to express is only a wish of the human mind to achieve the vanishing limit of the thickness of the boundary. I find it trivial to note that this limit can not be achieved in practice i.e. the wish can not be fulfilled. This is because, if the boundary vanishes then it can not be seen and the concept of area can not be expressed. Therefore, the truth of the area is valid only in a limiting sense that I can consider complete by assumption. This assumption is the ignorance of the context which is here the thickness of the boundary. Then, the law of excluded middle is the limiting case of the whole truth that includes the context where the limit is achieved by the ignorance of the context i.e. any judgment like ``true'' or ``false'' are only limited by the ignorance of the context.

The logician's attitude to achieve completeness of logic gets reflected more profoundly from the next statement of Frege that I quote as follows.   On page no. 161 of ref.\cite{fregedef}, Frege writes: \vspace{0.1cm}

{\small ``{\it It may be difficult to satisfy the demands of logic always in giving definitions; but it must be possible.}''}\vspace{0.1cm}

The statement is just a reflection of the attitude of a logician to achieve completeness of logic. This is because the phrase -- ``must be possible'' -- is a forceful assumption that results from a hard coded belief. Any human mind blinded by any belief only chases the same, even if it has to ignore in the process any immediately perceivable belief-defying truth. Frege's case is an example where the logician chases completion of logic by believing in the same and ignores the impossibility of attaining such feat that is directly realizable through his own metaphor.  


Now, the question arises whether it is at all possible to make formal analysis if a concept can never be defined sharply as Frege asserts on page no. 166 of ref.\cite{fregedef}:\vspace{0.1cm} 

{\small ``{\it ....without complete and final definitions, we have no firm ground under-foot, we are not sure about the validity of our theorems, and we cannot confidently apply the laws of logic, which certainly presuppose that concepts, and relations too, have sharp boundaries.}''}\vspace{0.1cm}

I believe it is a question of attitude. For me it is easy to admit that logical analysis can certainly be done but only with concepts whose definitions are considered as complete by assumption and not through achievement by means of any mode of expression.  I find it too easy to realize through an immediate self-inquiry, that is, I can never answer the question -- ``Who/what am I?'' -- through any mode of expression that I can access. In fact, I get confused whether to write ``who'' or ``what'' because it is the ``I'' that makes the difference. In the language of the logician, I can never completely define the concept of ``I'' through any mode of expression and I find any truth that I define and I claim to define is readily incomplete due to my inability of expressing the concept that I symbolize as ``I''. Such an attitude of self-inquiry, if adopted by the logician, he will find it easy to realize that completeness of any expression is achievable only by assumption. 


     	\subsection{The logician's silence: a choice} 
   While discussing about ``definition'' it is worth mentioning the involved writings of Quine in refs.\cite{quine,quine1,quine2,quine3}. However, I have no intention of presenting a detailed review of different views of various logicians on ``definition'' in this article because that is unnecessary in the present context. Rather the aim is to discuss certain basic logico-linguistic issues regarding physics which the logician does not care about. As I have pointed out in ref.\cite{ll}, the logician never dares to ask the question that how can the physicist claim to be logical while using the same symbol to convey different meanings, viz. number and not number. In fact, the literature reveals an opposite flow i.e. the physicist, based on the empirical essence of physics, has questioned the foundations of logic \cite{vonneuman, finkelstein, putnam, reichenbach, dummett}. Therefore, the logician may consider the present work as the foothold to question the physicist, regarding the issues concerning the logico-linguistic foundations of physics (or the expressions of the physicist), which the concerned authors of ref.\cite{vonneuman, finkelstein, putnam, reichenbach, dummett} take for granted in the process. Although I shall provide shortly a glimpse of how a simple logico-linguistic inquiry, by the logician, into the foundations of calculus could have unfounded the cornerstone on which the authors of ref.\cite{vonneuman, finkelstein, putnam, reichenbach, dummett,quantumlogic} have built upon, let me first spend a few more words on the attitude of the logician towards physics.
       
     \subsubsection{Quine on physics}
     To reflect the logician's attitude towards physics and the physicist, I analyze some of Quine's relevant statements from ref.\cite{quine1} as examples.     
     
     Quine writes on page 18-19:
     \vspace{0.1cm} 
     
     {\small``{\it But what of physics? An antinomy arose between the undular and the corpuscular accounts of light; and if this was not as out-and-out a contradiction as Russell's paradox, I suspect that the reason is that physics is not as out-and-out as mathematics.}''} \vspace{0.1cm}
     
     Quine's suspicion is correct but he does not point out what distinguishes physics from mathematics. The answer is -- physical dimension. And, there is no ``out and out'' answer to the question whether physical dimension is number or not number. If Quine would have realized this, he would have understood the essence of the middle way. Rather, without such realization, he takes the physicist and his physics for granted and only draws an analogy like the following: 
     \vspace{0.1cm}
     
     {\small ``{\it Again, the second great modern crisis in the foundations of mathematics -- precipitated in 1931 by Godel's proof [2] that there are bound to be undecidable statements in arithmetic -- has its companion piece in physics in Heisenberg's indeterminacy principle.}''.}
     \vspace{0.1cm}
     
     Unfortunately such restricted views on mere analogies do not lead to any useful contribution from the logician that can be of help to the physicist. Ironically, it is the Heisenberg uncertainty (``indeterminacy'') principle\cite{heisenberg} that lies at the foundation of ref.\cite{vonneuman} that challenges the foundations of logic from which the logician earns his living. Strangely, the logician remains silent even though the physicist does not follow logic (in the sense of the logician) while making use of physical dimension to question the foundations of logic. Why? I believe that the failure to understand the essence of the middle way leads the logician to believe in a possibly achievable completeness of logic and since any truth is only relational, the truth of logic becomes the ultimate reality for the logician. Since the logician does not realize the emptiness of logic, he is unable to question the physicist. Such shortcoming in the logician's strength of reasoning reflects even more properly from Quine's words as he writes on page no. 42:
     \vspace{0.1cm}
     
     {\small ``{\it The totality of our so-called knowledge or beliefs, from the most casual matters of geography and history to the profoundest laws of atomic physics or even of pure mathematics and logic, is a man-made fabric which impinges on experience only along the edges. }''} \vspace{0.1cm}

  Here, Quine distinguishes between physics and pure mathematics. Then I wonder why the adjective ``pure'' is necessary. Does that mean there is some impure mathematics as well? If there is, then how is that different from pure mathematics? Is it that he wants to refer to physics as impure mathematics? If he does, then what is that impurity? Is it physical dimension? 
  
  It is interesting to note that Quine uses the words ``physical lexicon'' on page no. 99 of ref.\cite{quine2}:\vspace{0.1cm}
  
  {\small``{\it But the mathematical and logical component, purified of all physical lexicon, is not called physics.}''}\vspace{0.1cm}
    
  where he discusses his views on the categorical divisions between sciences viz. logic, mathematics and physics. Then I wonder why he does not inquire how physical lexicon can be treated as number by the physicist during dimensional analysis while it is a priori not number simply because it distinguishes physics from mathematics. Only Quine knows. But, can the logician answer?

     At the end of the day it becomes a matter of personal choice, albeit limited by the ability of the logician, what he wants to question i.e. the logician has the right to choose whether he wants to remain reverent towards the physicist or he wants to make an inquiry regarding the illogical structure of the physicist's expressions.   
     
    \subsection{The logician's missed opportunity: an example}\label{missedopp}
    The fact that the logician simply chooses what to question becomes very clear to me when I find that he, in spite of being the master of nitpicking with words (which is necessary for formalizing language), does not write a single word on why Cauchy's verbal definition of derivative is not converted properly into mathematical expressions. I explain this by quoting Cauchy from 
    page no. 11 of ref.\cite{cauchycal} to note how he defined, what is known today as, ``derivative of a function'' -- \vspace{0.1cm}

{\small ``{\it $\cdots$ function $y = f (x)\cdots$ variable $x\cdots$ an infinitely small increment attributed to the variable produces an infinitely small increment of the function itself. $\cdots$ set $\Delta x = i$, the two terms of the ratio of differences
     	\begin{eqnarray}
     		\frac{\Delta y}{\Delta x}=\frac{f(x+i)-f(x)}{i}.\label{cder}
     	\end{eqnarray}	
     	will be infinitely small quantities. $\cdots$ these \underline{two terms indefinitely and simultaneously will approach} \underline{the limit of zero}, the ratio itself may be able to converge
     	toward another limit,...}''}\vspace{0.1cm} 

     The mathematical expressions corresponding to Cauchy's ``infinitely small quantities'', in modern notation, look like ``$\Delta x\to 0,\Delta y\to 0$'' in the above quoted example. The point to be noted here is that,  according to Cauchy's prescription, {\color{blue}{\it one needs both ``$\Delta y\to 0$'' and ``$\Delta x\to 0$'' to be satisfied to define the derivative}}. Therefore, 
     expression (\ref{cder}) 
     looks like the following:
     \begin{align}
     	\boxed{\frac{dy}{dx}:=\lim\limits_{\substack{\Delta y\to 0\\ \Delta x\to 0}}\frac{\Delta y}{\Delta x}=\frac{f(x+\Delta x)-f(x)}{\Delta x}\qquad\ni y=f(x).\label{cderp}}
     \end{align}
     
    Here, the symbol ``$\ni$'' stands for ``such that'' and ``$:=$'' stands for ``defined as'' and I have skipped the unnecessary step of setting $\Delta x=i$. Even if I keep aside the question regarding the use of the word ``quantity'', I think it can be directly observed how the conversion of the verbal statements into the corresponding mathematical expression leads to a different scenario than what is taught generally in textbooks -- the concept of limit is applied to the denominator only  e.g. see ref.\cite{rudin,apostol,spivak,ross}. No sophisticated formalized language, like what the logicians write, is needed to understand and demonstrate this. Yet, needless to say, the implication is far reaching. Before providing a glimpse of such an implication, I  discuss how the logician only shows a reverential attitude towards the physicist rather than making any logico-linguistic inquiry regarding the truthfulness of his expressions.

    \subsubsection{A logico-linguistic inquiry into the language of the logician}
     In {\it The Ways of Paradox}, which is the first essay in ref.\cite{quine3}, Quine makes a logico-linguistic analysis of several paradoxes which necessitates very precise nitpicking of words. However, in the relevant passages regarding calculus in {\it The Foundations of Mathematics} in ref.\cite{quine3}, such preciseness of the inquiry goes missing. I explain this by quoting the relevant passages and by offering suitable remarks as follows.
     
   Quine writes:\vspace{0.1cm} 
        	
     	 ``{\small {\it It was the idea of a fractional quantity infinitely close to zero, yet different from zero. It seemed to be needed in the study of rates, which was the business of the differential calculus.}}''\vspace{0.1cm}

         I note that Quine, who distinguishes between mathematics and physics due to the use of ``physical lexicon'' (in his own words) in the later, does not notice that quantity and number differ due to the involvement of the ``physical lexicon''. It is the comparison of two quantities that involves some number\cite{ll}. Then, I wonder how Quine compares quantity and number. For example, if I question: ``{\it Which one of the following statements is meaningful? -- (a) The mass of the object is $1$ gram. (b) The mass of the object is $1$.}'',  I believe that any reasonable thinker, free from the influence of any dogma or authority, will agree with me that only option {\it (a)} is meaningful. And then the associated number depends on the unit (i.e. the chosen standard quantity) with which the concerned quantity is being compared. It is due to this reason ``quantity infinitely close to zero'' is also meaningless. Rather, what seems meaningful to me is a statement like ``quantity infinitely small compared to the unit''. It is quite surprising to me why the logician, who is a fine nitpicker of words, fails to see such a basic issue. The same ignorance of this logico-linguistic issue continues to reflect through subsequent writings of Quine while he uses verbal expressions like ``distance.....is zero'', ``infinitesimal distances'', ``infinitesimal time'', etc. I wonder why the logician fails to see such a vivid fact that distance is not number. So, how can it be ``zero''? If distance or time is ``infinitesimal'', then why does the logician NOT raise the question ``with respect to what unit of length or time?'' 
         
         Quine continues to write:\vspace{0.1cm}


  ``{\small {\it Still the resulting calculus proved indispensable in reasoning mathematically about rates.}}''\vspace{0.1cm}

As far as I understand, reasoning is primarily done verbally. To transmit the arguments truthfully to the mathematical jargon it is necessary to be careful about the conversion principles (which the logician is a master of). Then, how do the ``physical lexicons'' go missing from the language of the logician? The question arises from Quine's own distinction between mathematics and physics on the basis of  ``physical lexicons''. I believe, it is a simple fact to observe that mathematics, being devoid of ``physical lexicons'' can not be used to convert the words like ``distance'', ``time'', etc. Then, is it that some {\it physico}-mathematical reasoning is required to incorporate ``physical lexicons''? And, if that is so, is not it that all the expressions concerning the word ``infinitesimal'' must involve statements that express comparison between two quantities? A logico-linguistic inquiry regarding the elements of the physicist's expressions becomes inevitable and necessary \cite{ll}. \vspace{0.1cm}
 
 However, the logician, instead of making such a logico-linguistic inquiry, shows a reverential attitude towards the physicist (or the mathematician who applies mathematics to do physics without being concerned about ``physical lexicons''). Such reverence is manifest as Quine writes further:\vspace{0.1cm}

       {\small ``{\it  Cauchy and his followers in the nineteenth century solved the problem....... Each such distance-to-time ratio will multiply out to about a mile a minute, if the time interval is short.}''}\vspace{0.1cm}

       Here, some simple questions can be raised regarding the term ``ratio''. In arithmetic, I have come across ratio of two numbers and also in physics, I have learned the lesson that only quantities with the same physical dimension (``lexicon'') can be compared. Then, what is the meaning of ``distance-to-time ratio''? Is not it that one needs to specify some arithmetical rules for the use of ``physical lexicons'' to give meaning to such a phrase? But, how can one specify arithmetic rules for ``physical lexicons'' because, as Quine writes, mathematics is devoid of ``physical lexicons''. Why does the logician ignore such vivid logico-linguistic questions?

    \subsubsection{A glimpse of the impact of direct reasoning through logico-linguistic inquiry: towards Heisenberg uncertainty principle from Cauchy's definition of derivative}\label{uncertainty}
The physicist has challenged, at times, the foundations of logic based on what is known today as ``quantum mechanics'' and the most distinct feature of this theory that plays a pivotal role in the physicist's arguments is the  Heisenberg uncertainty principle\cite{heisenberg}. While today such logic is known as ``quantum logic''\cite{quantumlogic}, it started with Birkhoff and von Neumann who called it ``The logic of quantum mechanics''\cite{vonneuman}, Finkelstein coined the term ``Physical logic'' \cite{finkelstein} and Putnam called it ``Empirical logic'' \cite{putnam}. There are others like Reichenbach\cite{reichenbach} and Dummett \cite{dummett} who also proceeded along such lines of thought. Now, I provide a glimpse of how the logician can make a direct logico-linguistic inquiry into the foundations of physics to show the physicist that the Heisenberg uncertainty principle, based on which the physicist builds his arguments regarding quantum/physical/empirical logic, follows directly from the criteria for the definition of derivative. 
   
   To apply Cauchy's definition of derivative to define instantaneous velocity, let me denote the units of length and time as $\lambda_0$ and $T_0$ respectively. Then, I write the change in position of an object as $\Delta x=\Delta n_x\lambda_0$ and the time interval as $\Delta t=\Delta n_tT_0$, where $\Delta n_x, \Delta n_t$ are real positive numbers. So, I can write an inter-conversion among  verbal expressions and physico-mathematical expressions as follows:
   \begin{eqnarray}
   \text{\footnotesize ``$\Delta x$ is infinitesimally small compared to the length unit $\lambda_0$''}&\Leftrightarrow&\Delta x\ll\lambda_0~~\Leftrightarrow~~\Delta n_x\to 0,~~\label{delx}\\
    \text{\footnotesize ``$\Delta t$ is infinitesimally small compared to the time unit $T_0$''}&\Leftrightarrow&\Delta t\ll T_0~~\Leftrightarrow~~\Delta n_t\to 0,~~\label{delt}
   \end{eqnarray}  
   where the expressions $\Delta n_x\to 0, \Delta n_t\to 0$ corresponds to the limits. I prefer to write ``$\Delta n_x\lll 1, \Delta n_t\lll 1$'' instead of ``$\Delta n_x\to 0, \Delta n_t\to 0$'' considering what I have discussed in ref.\cite{ll} regarding ``zero quantity''. However, I continue to use the notation so that the reader can connect with the usual convention. 
   
   Now, under the validity of such conditions, instantaneous velocity can be written in terms of physico-mathematical expressions using Cauchy's definition of derivative as follows:
   \begin{eqnarray}
   	\boxed{v\equiv \frac{dx}{dt}:=\lim\limits_{\substack{\Delta n_x\to 0\\ \Delta n_t\to 0}}\frac{\Delta x}{\Delta t}=\(\lim\limits_{\substack{\Delta n_x\to 0\\ \Delta n_t\to 0}}\frac{\Delta n_x}{\Delta n_t}\)\frac{\lambda_0}{T_0}=n_v v_0 ~\ni  n_v:=\lim\limits_{\substack{\Delta n_x\to 0\\ \Delta n_t\to 0}}\frac{\Delta n_x}{\Delta n_t}~\& ~v_0:=\frac{\lambda_0}{T_0}.}
   \end{eqnarray}

   Now, to write down Newton's second law, the derivative of momentum needs to be defined according to Cauchy's prescription. So, let me consider momentum $(p)$ to be a quantity in its own right. I denote a change in $p$ as  $\Delta p=\Delta n_p p_0$ where $p_0$ is the chosen unit of momentum. Then the instantaneous rate of change of momentum can only be defined when the following conditions are fulfilled:
   \begin{eqnarray}
\text{\footnotesize``$\Delta x$ is infinitesimally small compared to the length unit $\lambda_0$''}&\Leftrightarrow&~\Delta p\ll p_0~\Leftrightarrow~\Delta n_p\to 0,\label{delp}\\
\text{\footnotesize``$\Delta x$ is infinitesimally small compared to the length unit $\lambda_0$''}&\Leftrightarrow&~\Delta t\ll T_0~\Leftrightarrow~\Delta n_t\to 0.
  \end{eqnarray}
   Then, the definition of instantaneous rate of change of momentum, according to Cauchy's prescription, looks like the following:
   \begin{eqnarray}
   	\boxed{ \frac{dp}{dt}:=\lim\limits_{\substack{\Delta n_p\to 0\\ \Delta n_t\to 0}}\frac{\Delta p}{\Delta t}=\(\lim\limits_{\substack{\Delta n_p\to 0\\ \Delta n_t\to 0}}\frac{\Delta n_p}{\Delta n_t}\)\frac{p_0}{T_0}=n_F F_0 ~\ni  n_F:=\lim\limits_{\substack{\Delta n_p\to 0\\ \Delta n_t\to 0}}\frac{\Delta n_p}{\Delta n_t}~\& ~F_0:=\frac{p_0}{T_0}.}
   \end{eqnarray}
   
   Since the conditions (\ref{delx}) and (\ref{delp}) are necessary for writing down classical mechanics, then a I can certainly write down the following derived condition:  
   \begin{eqnarray}
   	\Delta x.\Delta p\ll L_0\quad\ni L_0:=p_0\lambda_0.
   	\label{cmech}
   \end{eqnarray}
   Therefore, if the above condition is violated then the basic definitions of the quantities (discussed above) which are required to do classical mechanics, do not hold anymore. This violation condition can be written as follows:  
   \begin{eqnarray}
   	\boxed{\Delta x.\Delta p\gtrsim  L_0.}\label{qmech}
   \end{eqnarray}
Certainly $L_0$ is restricted by our ability to chose the units of measurement and hence a more precise reasoning is required to argue how $L_0$ is equivalent to Planck's constant that appear in Heisenberg uncertainty principle\cite{heisenberg}, which I intend to present elsewhere. Nevertheless, I believe, I have been able to present a glimpse of how direct reasoning can also be effective without the necessity of formalization. The logician, if makes such logico-linguistic inquiry into the foundations of physics, then it will improve science as a whole. 



Unfortunately, this is not the case in general. For example, Quine only mentions Heisenberg uncertainty (``indeterminacy'') principle with reverence and without further investigation in terms of logic and language. 
I consider this as a missed opportunity of the logician to impact science, as a whole, with simple and direct reasoning. Although I do not deny that the logician's formalized language is required, however, I also can not get unnecessarily detached from direct and simple logico-linguistic inquiry that can lead to efficient reasoning for the betterment of science.

      }

\section{Comments for the philosopher}\label{philosophers}

As far as philosophy of physics is concerned I find it convenient to categorize the philosopher into two classes -- one that actively contributes to physics without being worried much about the structure of the language e.g. the authors of ref.\cite{heisenbergphil,boltzmannphil,duhem} and the other that does not contribute to physics but concentrates on the structure of language used by the physicist, etc. e.g. the authors of ref.\cite{nagel,popper,wittgenstein,carnap,kuhn}. To make the difference between the two classes clearer, I quote Maxwell from page no. 41 of ref.\cite{maxwell}:\vspace{0.1cm}

{\small ``{\it The torsion-balance was devised by Michell for the determination of the force of gravitation between small bodies, and was used by Cavendish for this purpose. Coulomb, working independently of these philosophers, reinvented it, and successfully applied it to discover the laws of electric and magnetic forces; and the torsion-balance has ever since been used in all researches where small forces have to be measured. }''}\vspace{0.1cm}
	
The philosopher of the present day (e.g. the authors of ref.\cite{popper,wittgenstein,nagel,carnap}), whom I call ``the philosopher'' henceforth, is not like those who Maxwell called ``philosophers''. Maxwell's philosophers (also include the authors of ref.\cite{boltzmannphil,duhem,heisenbergphil}) actively contributed to the progress of physics. Today, philosophy of physics\footnote{I avoid using ``philosophy of science'' and rather choose to use ``philosophy of physics'' to keep my viewpoint attached to the present context i.e. an inquiry regarding the foundations of physics.}, hence the philosopher of physics, does not affect physics directly, if not at all. As it appears to me, a possible reason is the detachment of the philosopher from practical essence of doing physics and attachment with abstraction in accord with the theoretical physicist of modern days. Consequently the philosopher, instead of making an inquiry regarding the language of the physicist, adopts the language of the physicist by taking it for granted and then plays with the structure of such language. As a result, if the physicist is in trouble, so is the philosopher -- no progress. The philosopher does not show a path, not even theoretically, which the physicist can not see to find a solution to some problem. This gives the chance to the physicist to blame the philosopher of physics regarding his contribution to physics e.g. on page 135 of ref.\cite{hawking}, Hawking writes: \vspace{0.1cm}

{\small ``{\it ......the people whose business it is to ask why -- the philosophers -- have not been able to keep up with the advance of scientific theories.}''}
\vspace{0.1cm}

 I must refrain from making such judgments about the philosopher, unlike the physicist, as I find the above quote quite disrespectful. Rather I prefer to indulge in a discussion regarding how the philosopher can take it to the physicist by asking some painful ``why''-s and shake the cherished foundations on which the physicist builds his dogma. Since I am concerned about how truthfully the physicist converts his experiences into the corresponding expressions while being operational, I put forward a question regarding a very basic statement adopted by the physicist while writing down theories, which the philosopher can immediately appreciate without having to ``keep up'' with the technicalities which the physicist dwell on in modern days e.g. see ref.\cite{hawellis}.

 \subsection{Why is there a time lapse to realize ``$t=0$''?}\label{t0}
 In physics, one finds no problem in writing the statement ``$t=0$''. Certainly it is meaningless writing because of dimensional mismatch, as it goes for any other quantity like length, mass, charge, etc\cite{ll}. Neither the physicist himself, nor the philosopher (nor the logician, especially the logical empiricist\cite{logicalempiricism1,logicalempiricism2,logicalempiricism3}) finds any trouble in accepting such a statement (if the philosopher realizes the problem at all). Even if I rectify the statement by saying that ``$t=0$'' means ``zero unit of time'', there is a clear defiance of experience that is reflected through such a statement. To experience that ``zero unit of time'' can not be experienced in practice, I do not need to ``keep up with the advance of scientific theories''. It is a matter of self-inquiry through which I can ask myself, `` Am I being truthful in expressing my perception when I write ``zero unit of time''?'' The answer that I get is ``no'' because whatever clock I use to make any measurement of time, the initiation of the running of the clock takes some time that can not be captured within the time measuring process of the clock. That is, the underlying mechanism that lets the clock be a ``clock'' takes some time to initiate the process through which the clock maintains its timing.
 
 Ironically, it is the word ``experience'' that Einstein uses repeatedly in formulating the foundations of his relativity theories in the introductory sections of ref.\cite{einsr}, although he does not explain how to justify the experience of ``$t=0$''. As a matter of fact, ``$t=0$'' is an accepted statement used to explain ``initial condition or the beginning'' of some phenomenon in theoretical physics. But, any honest experimental physicist will agree that ``zero unit of time'' can not be experienced -- there is always an experience of time lapse, however negligible it may be compared to the chosen time unit, in the context of the observation, when an observation of some phenomenon is initiated. So, as it stands, the statement ``$t=0$'' is only helpful for doing calculations of theoretical physics, but it defies immediate experience. Such a statement can be tolerated, albeit with a bit of skepticism, as long as it is used to write theories that explain observed phenomena. However, the problem begins when the physicist builds his dogma on the basis of such a statement irrespective of whether his statement has any empirical basis at all. Such is the case when the physicist indulges in explaining the ``origin of universe'' and discusses about ``a complete theory'' that can explain ``everything'', without realizing that he has not even been able to express the most basic of his perceptions regarding ``zero unit of time''\cite{hawking}. Can the philosopher hold enough courage to ask the physicist why such attitude itself is even permissible while doing physics in the pursuit of truth? If the physicist's investigations are based on the theories of relativity, which can not take account of such an elementary bit of experience, how can the physicist (who is a part of the universe) claim to be truthful at all while seeking to find the truth regarding ``the universe''?  
 

\subsection{The philosopher's choice: primacy of sophistication or truthfulness}
For me, philosophy has no barrier -- rather it is a way to unshackle the mind from any limitations of thought and bodily action. Therefore, I consider it as a loss of freedom when I limit my thought because of some other person's dogma. However, there is also the freedom of choice. The philosopher can always choose not to question the dogma of the physicist and just ignore crucial ``why''-s like the one I discussed in the previous subsection. In that case, such philosophy seems to be useless in the pursuit of truth and Hawking's complain against the philosopher stands firm, irrespective of the number of  pages the philosopher fills or what specific jargon and sophisticated notations he uses to express his thought. Feynman may promptly say that ``{\it Scientists are explorers, philosophers are tourists.}'' (see page no. 260 of ref.\cite{feynman}). 

Rather, Maxwell's ``philosophers'' seem to be more useful from this perspective as he is directly attached to experience of ``$t=0$'' through experiment. So, with Maxwell's ``philosophers'', a nitpicking of words of their experience (like what has been done regarding the definition of electric field in ref.\cite{ll}) is more likely to have immediate effect on physics, and hence will be of guidance to the physicist. If the philosopher follows the footsteps of Maxwell's ``philosophers'', then he can guide the physicist rather than having to ``keep up'' with ``the advance''. Feynman may then call them ``explorers''. To do this, the philosopher needs to explore the truthfulness of the relation between experience and  expression, rather than focusing on the sophistication of the language which the physicist already has used to express his experience. Indeed, it is in such spirit that gets manifested in ref.\cite{ll}.

Digressing a bit, albeit necessarily due to the context, I may provide the following remarks regarding ref.\cite{ll}. It is due to above reasons, if the reader categorizes ref.\cite{ll} as a work on philosophy of physics, then certainly it inherits a distinctive feature when compared to other works on philosophy of physics and that is -- questioning, rather than adopting, the language of the physicist with an aim to directly address certain the truthfulness of the relation between experience and expression of the physicist. It is directed towards sharpening the language of operations rather than the language of axioms i.e. the truthfulness of the expression of experience gets the priority over the sophistication of language. And, it is due to this single most reason, that ref.\cite{ll} is best presented in a form that is devoid of unnecessary abstract notations and specific jargon which do not serve any purpose in resolving the issues concerning the foundations of physics i.e. the elements of the physicist's expressions of experience. The skeptic philosopher can still choose sophistication as the primary motive, but I choose the primacy of truthfulness based on self-inquiry. If the philosopher is still not convinced, I may invite him for a scholarly debate on the following crucial ``why'': {\it Why is there a visible dot to represent ``$r=0$'' or ``zero length''?}\cite{essay} 
 
 \section{Outlook}
The physicist needs the logician and the philosopher. But, questions remain. Can the logician bring out his sharpness of reasoning to question the physicist? Can the philosopher delve into the infinitude of immediately realizable truths to provide more truthful means to express the experience to guide the physicist on the intellectual ground? However, then the categories of these disciplines become questionable. Can the logician, the philosopher and the physicist look beyond their own categories, and respective sophistication, to find the emptiness of each category? Certainly, I believe, these questions are worth considering. If the singularity problem in physics can be resolved by a truthful conversion of the experience of seeing a dot on the paper\cite{essay}, if the essence of  middle, which is excluded by choice by the logician, can be understood by drawing a closed line with a pencil on the paper\cite{ll}, if the Heisenberg uncertainty principle can be realized in few steps from Cauchy's definition of derivative, then why is it so hard for the logician, the philosopher and the physicist to break free of the shackles of categories and become `nobody' for the betterment of science as a whole?
 
     

{\it Acknowledgment:}  This work is supported by the Department of Science and Technology of India through the INSPIRE Faculty Fellowship, Grant no.- IFA18-PH208.

\end{document}